\documentclass[twocolumn,amsmath,amssymb]{revtex4}

\usepackage{graphicx}
\usepackage{dcolumn}
\usepackage{bm}
\newcommand{\etal}{\emph{et al.}}
\newcommand{\be}{\begin{equation}}
\newcommand{\ee}{\end{equation}}
\newcommand{\bfig}{\begin{figure}}
\newcommand{\efig}{\end{figure}}
\newcommand{\incl}{\includegraphics}

\begin{document}      

\title{Large enhancement of the thermopower in Na$_x$CoO$_2$ at high Na doping}
\author{Minhyea Lee$^1$, Liliana Viciu$^2$, Lu Li$^1$, Yayu Wang$^{1,\dagger}$, M. L. Foo$^2$, 
S. Watauchi$^2$, R. A. Pascal Jr.$^2$, R. J. Cava$^2$ and N. P. Ong$^1$
}
\affiliation{
Departments of Physics$^1$ and Chemistry$^2$, Princeton University, 
Princeton, New Jersey 08544, U.S.A.\footnote{Nature Materials, \emph{in press}}
}

\date{\today}      

\maketitle                   
{\bf Research on the oxide perovskites has uncovered electronic properties 
that are strikingly enhanced compared with those in conventional metals.  
Examples are the high critical temperatures
of the cuprate superconductors and the colossal magnetoresistance  
in the manganites.  The conducting layered cobaltate $\rm Na_xCoO_2$ 
displays several interesting electronic phases as $x$ is varied~\cite{Terasaki,Wang,Foo}, 
including water-induced superconductivity~\cite{Tanaka} and an insulating state~\cite{Foo} 
that is destroyed by field~\cite{Balicas}.  Initial measurements~\cite{Terasaki} showed that, 
in the as-grown composition, $\rm Na_xCoO_2$ 
displays moderately large thermopower $S$ and conductivity $\sigma$.
However, the prospects for thermoelectric cooling applications faded 
when the figure of merit $Z$ was found to be small at this 
composition (0.6$<x<$0.7).  Here we report that, in the poorly-explored
high-doping region $x>$0.75, $S$ undergoes an even steeper enhancement.
At the critical doping $x_p\sim$ 0.85, $Z$ (at 80 K) reaches 
values $\sim$40 times larger than in the as-grown crystals.  We discuss
prospects for low-temperature thermoelectric applications.
}

In the large-$x$ region of $\rm Na_xCoO_2$ ($x>$0.75), progress has been hampered
by difficulties in growing single crystals as well as by phase-separation effects
which appear above $x_p$.
Powder neutron diffraction~\cite{Huang} has revealed subtle shifts in the 
Na ions as $x$ is increased above 0.75.  The doping interval $0.75<x<x_p$ 
is a homogeneous phase $H_2$, in which the thickness $t$ of 
the $\rm CoO_2$ layers undergoes a steep increase~\cite{Huang}.  Above $x_p$,
the neutron results indicate phase separation (the mixed-phase region is labelled $H_2$+$H_3$).  
Muon spin rotation~\cite{Mendels05}, nuclear magnetic resonance (NMR)~\cite{deVaulx} and 
susceptibility~\cite{Lang05,deVaulx} experiments also suggest mixed phases in highly-doped samples.
Finally, the limiting phase $H_3$ ($x\rightarrow 1$) is inferred from NMR and 
magnetic susceptibility to be a non-magnetic insulator
with a slight admixture of a residual metallic phase~\cite{deVaulx,Lang05}.  

In a thermopower experiment, the current density ${\bf J}' = \sigma{\bf E}$
produced by the $E$-field is cancelled by the thermoelectric
current ${\bf J} = \alpha(-\nabla T)$ driven by the applied gradient $-\nabla T$
($\alpha$ the thermoelectric or Peltier conductivity). 
Whereas the thermopower $S = \alpha/\sigma$ is the quantity usually reported, 
we have found that $\alpha$ provides a more incisive probe for sorting out the 
results in the region $H_2+H_3$.  Results from 10 crystals (of nominal 
size 400$\times$700 $\mu$m) with $x$ ranging from 0.71 to $\sim$1 reveal that $\alpha$ rises  
steeply to a peak at the threshold doping $x_p$, with $\sigma$ nearly constant
(as shown in Fig. \ref{rho}, samples are labeled 1-10 in order of increasing $x$).  In
the mixed region above $x_p$ both quantities decrease rapidly.

Crystals of $\rm Na_xCoO_2$ grow as either bilayer or tri-layer structures
(2 and 3 $\rm CoO_2$ layers per unit cell, repectively)~\cite{Viciu}.
The in-plane transport quantities are very similar in the 2 structures.  
However, at large $x$, the 3-layer crystals (Samples 2*, 9*, 10*) tend to lock to 
the commensurate dopings $x = \frac34$ and $\sim$1, which provide valuable checks.  

\bfig
\incl[width=7cm]{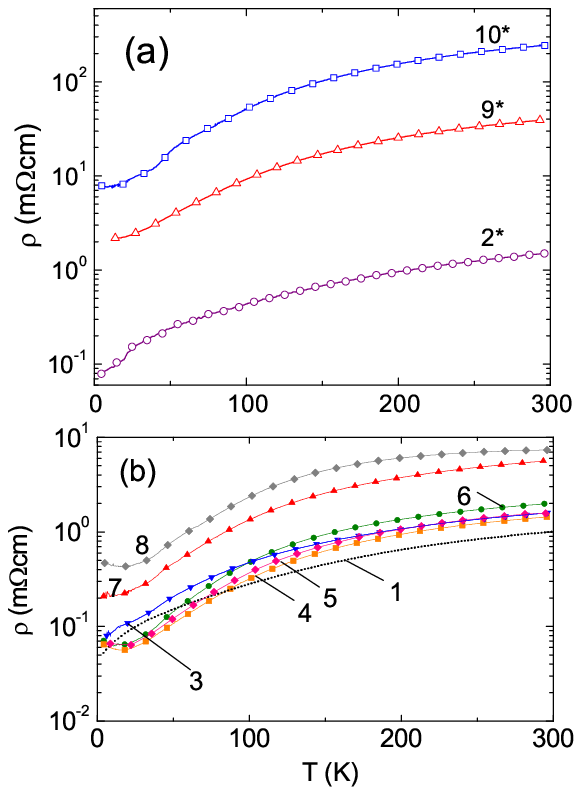}
\caption{\label{rho} In-plane resistivity $\rho$ vs. $T$ in the 3-layer (Panel a) 
and 2-layer (Panel b) cobaltate in log-linear scale.  The samples, numbered 
in order of increasing $x$, are: Sample 1 ($x\simeq$ 0.71), 2* (0.75), 3 (0.80), 
4 (0.85), 5 (0.88), 6 (0.89), 7 (0.96), 8 (0.97), 9* (0.99) and 10* (1.0).  
Asterisks indicate the 3-layer crystals (Panel a).  From $x$ = 0.71 to $x_p\sim$
0.85, $\rho$ does not change significantly (Samples 1--5).  Inside the mixed-phase
region $H_2+H_3$, however, $\rho$ increases rapidly (6--10*).  The persistent 
metallic profile inside $H_2+H_3$ suggests that conducting 
layers are embedded in an insulating matrix.  Scaling of the curves
of $\rho$  are not as satisfactory as for $\alpha$ (see below).  Contacts with
contact-resistance 2-10 $\Omega$ were attached by lightly
abrading the crystal surface and attaching Au wires with Ag paint (Du Pont 4922N).  The size of 
contact pads (see Supplement) leads to a total
uncertainty in the absolute values of $\rho$ of $\pm 15\%$.
}
\efig

Figure \ref{rho} displays curves of the observed $\rho$ vs. $T$ (temperature) in the 3-layer (Panel a) 
and 2-layer (b) cobaltate.  Beginning with the lowest curves
(Samples 1-4 in Panel b), we note that the profiles of $\rho(T)$ are metallic with the residual
resistivity ratio RRR $\simeq$ 20, comparable to the as-grown crystals ($0.6< x<0.7$).  
Significantly, in Samples 2*--6 (regions $H_2$ and slightly
beyond $x_p$), the room-temperature values $\rho(300)$ lie in the range 1-2 m$\Omega$cm.
As $x$ crosses $x_p$ into $H_2+H_3$
(Samples 5--8), $\rho(300)$ rises steeply to $\sim$240 m$\Omega$cm (in 10*),
which far exceeds the Mott limit.  Surprisingly, the curves retain the ``metallic'' profile even as $x\rightarrow 1^{-}$.  
This implies that, for $x>x_p$, the carriers strongly segregate into a fraction $f$ of the 
layers.  The observed $\rho(T)$ reflects the intrinsic resistivity of these conducting layers 
inflated by the geometric factor $1/f$.  Moreover, the persistence of metallic 
behavior to 4 K implies that the layers remain electrically connected down to low $T$.  
Thermal activation of carriers over low barriers, frequently seen in phase-separated 
systems, is not observed.

Curves of the thermopower are displayed in Fig. \ref{SvsT}a.  The 
curve with the smallest $S$ (Sample 1) is typical of as-grown crystals~\cite{Terasaki,Wang}.  
Across the $H_2$ region (Samples 2*, 3, 4), the $S$-$T$ curves show an increasingly negative 
curvature that grows into a broad peak of magnitude 240 $\mu$V/K at $\sim$130 K in 4.
In the mixed-phase region, the peak value rises further to 300-350 $\mu$V/K (Samples 5--8),
before settling to 200 $\mu$V/K as $x\rightarrow 1$ (Samples 9*, 10*).  These large $S$ values 
far exceed any reasonable extrapolation of the Sommerfeld 
expression $S = (k_B/e)(k_BT/\epsilon_F)$ ($k_B$ is Boltzmann's constant
and $e$ the electron charge).  To match the observed 
$\partial S/\partial T\simeq$ 3 $\mu$V/K$^2$ at low $T$, we would need
the Fermi energy $\epsilon_F$ to equal $\sim$30 K, which is unphysically small. 
The unusually large $S$, coexisting with low resistivities ($\sim 100\;\mu\Omega$cm in 
Samples 4--6), challenges our understanding of how
strong correlation enhances the Peltier effect.  Approaches include the use
of the Heikes formula~\cite{Maekawa,Chaikin,Wang}.  

\bfig
\incl[width=7cm]{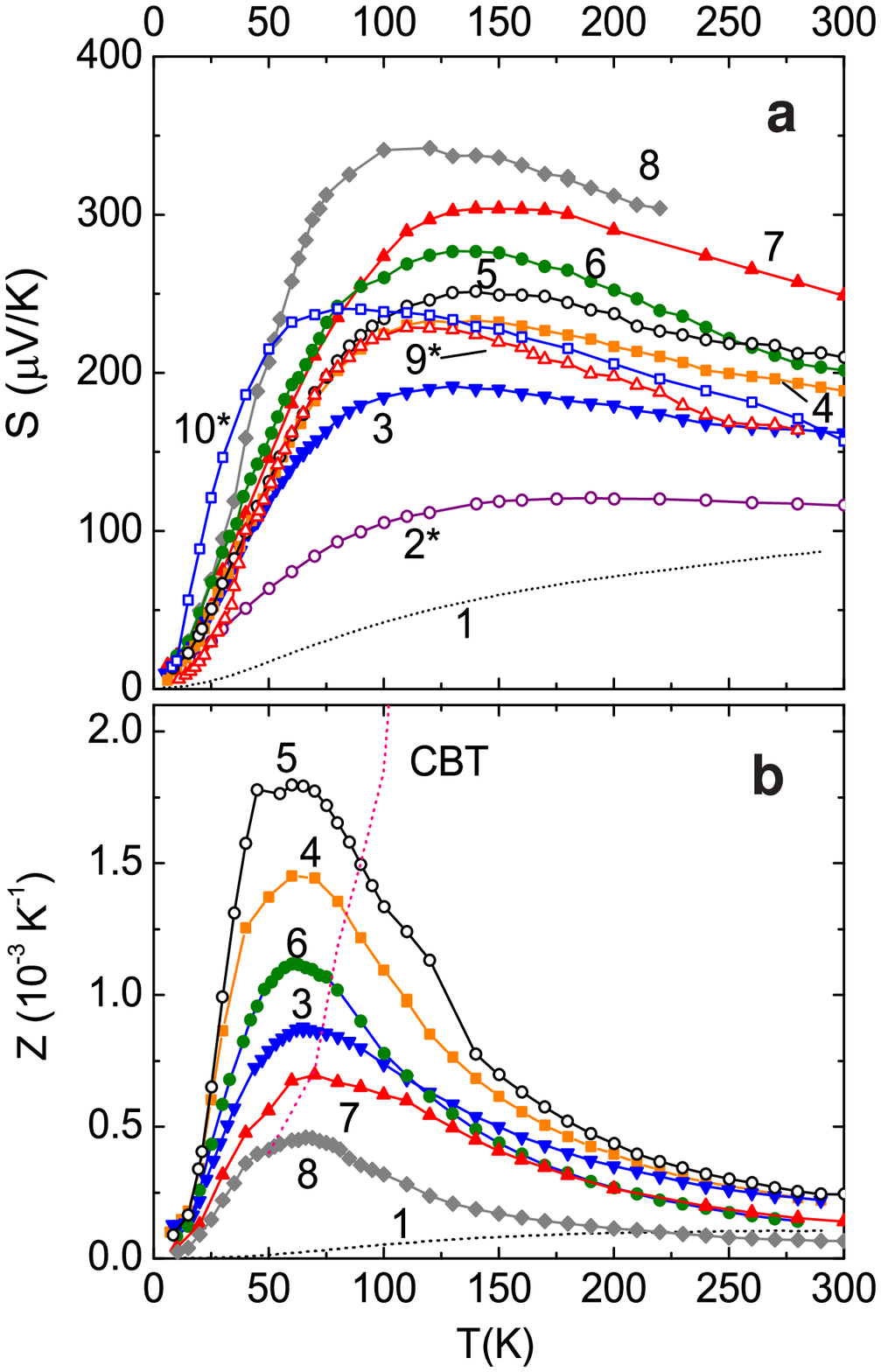}
\caption{\label{SvsT}  The in-plane thermopower $S$ (Panel a) and the 
figure of merit $Z$ (Panel b) in $\rm Na_xCoO_2$.  In Sample 1 (Panel a), 
$S$ is very similar to that of Terasaki \etal~\cite{Terasaki}.   As $x$ increases into
phase $H_2$ (2*--4), the profile of $S$ develops an
increasing bulge near 130 K that grows rapidly to peak values of 200-250 $\mu$V/K.
In the mixed-phase region, $S$ further increases to 300-350 $\mu$V/K (Samples 5--8),
before settling down to 228 $\mu$V/K in the limit $x\rightarrow 1$
(Samples 9*).  A striking pattern is that the $S$-$T$ profiles (all hole-type) are
nominally similar in shape in the mixed-phase region.  The exception is Sample 10*
in which $f\sim 1/200$ (see Supplementary Information).  Panel b shows curves 
of $Z = S^2/\rho\kappa$, with $\kappa$ measured separately (not shown).  
As $x$ increases from 0.71 (Sample 1), the peak value of $Z$
increases steeply to 1.8$\times 10^{-3}$ K$^{-1}$ in Sample 5. 
At higher $x$, the peak value of $Z$ falls rapidly because of the 
sharp increase in $\rho$.  At 80 K, $Z$ in Sample 5 is $\sim$40 times larger
than that in Sample 1. The dashed line labeled CBT is $Z$ reported~\cite{Chung} for $\rm CsBi_4Te_6$.
}
\efig

In Fig. \ref{SvsT}b, we display plots of the figure of merit $Z = S^2/\rho\kappa$ for some
of the samples ($\kappa$ is reported in Supplementary Information).  
Near the threshold $x_p$ (Samples 4, 5), $Z$ rises to a prominent maximum 
below 100 K.  As mentioned, the peak value of $Z$ at 80 K is $\sim$40 times larger than 
that in the as-grown composition (compare 5 with 1).  We discuss $Z$ below.
The dimensionless figure of merit $ZT$ is plotted in Supplementary Information.

The Peltier conductivity $\alpha$ shows nominally similar profiles (Fig. \ref{alpha}a).  
In Samples 1 and 2*, $\alpha$ displays broad peaks at 100 and 80 K, respectively, but remains fairly small. 
As we enter the $H_2$ region, however, $\alpha$ increases rapidly, displaying a sharply peaked 
profile in Sample 4.  In simple metals, $\alpha\sim (k_B^2/e)(T/\epsilon_F)\sigma$ 
decreases linearly to zero as $T\rightarrow$ 0.  
At high $T$, $\alpha$ tends to saturate to values in the range 1-10 A/mK.  Hence, the pronounced peak 
and large values of $\alpha$ reported here ($\alpha_{max}\sim$ 135 A/mK in Sample 4) 
are strikingly anomalous, and likely a consequence of strong correlation.

\bfig
\incl[width=7cm]{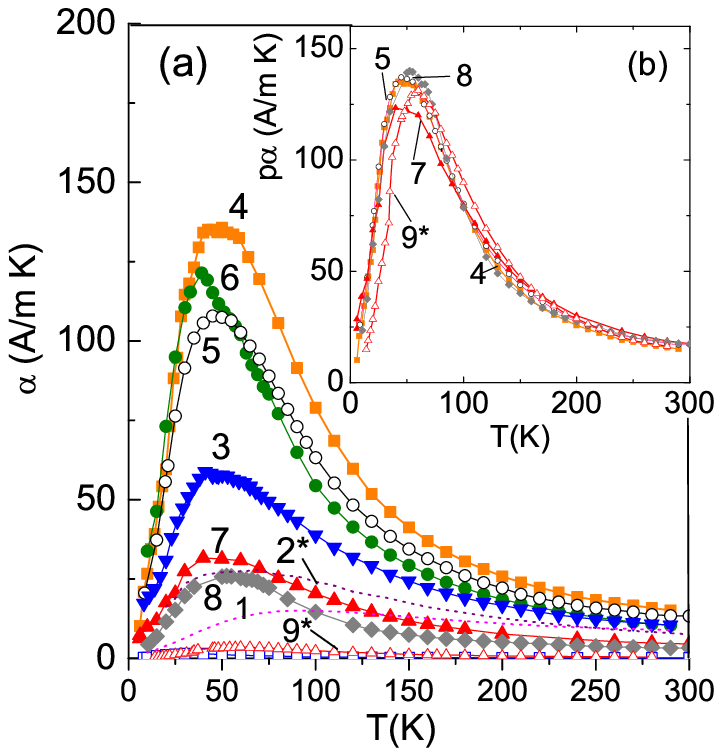}
\caption{\label{alpha} Curves of the Peltier conductivity
$\alpha$ vs. $T$ (Panel a) and scaling of the $\alpha$-$T$ curves in 
selected samples (Panel b).  As $x$ traverses the region $H_2$
(Samples 1--4) the peak in $\alpha(T)$ systematically shifts from
100 K to 50 K.  Because of the shift these curves cannot be scaled 
together.  However, in the region $H_2+H_3$ (5--10), 
the peak in $\alpha$ remains at 50 K.  Panel b shows that the  
curves of 5--10 match the template in 4 when multiplied by the scale
factor $p$ ($p$ = 1.3, 3.9, 5.4, 38 and 200 in Samples 5, 7, 8, 9*, and 10*, respectively).
Sample 6 has a cusp of unknown origin that makes it an exception.  Sample
10* is discussed in Supplement.
}
\efig

By tracking $T_{max}$ (the peak temperature of $\alpha$), we see that $\alpha$ vs. $T$
in 1--4 ($x\le x_p$) evolves differently from $\alpha$ in
5-10* (in $H_2+H_3$).  In the first group, $T_{max}$ decreases
from 110 K to 50 K (in 1 and 4, respectively).  The continuous
change in $T_{max}$ precludes a simple scaling analysis. 
By contrast, in the second group, $T_{max}$ remains fixed at $\sim$ 50 K.  Moreover,
by multiplying each curve by a scale factor $p$, we may match it to
the curve of 4 (Fig. \ref{alpha}b).  The exception is Sample 6 which  
has a cusp at 36 K (of unknown origin) that ruins the scaling.  Hence, as we 
traverse the homogeneous region $H_2$, the peak value of $\alpha$ increases rapidly
from 20 A/mK in 2* to 135 A/mK in 4, but the form of $\alpha$ vs. $T$
also changes continuously.  However, once we cross the threshold $x_p$
into the mixed phase, the profile is locked to that of 4.  

The simplest explanation of the scaling is that, in $H_2+H_3$, the carriers segregate 
into \emph{continuous} conducting layers embedded in a nearly insulating matrix.  The conducting
layers correspond to doping $x_p$ whereas the insulating matrix is at $x = 1.0$.  
The mean doping $x_{m}$ is then given by
\be
x_{m} = fx_p + (1-f), \quad\quad(x_p<x_{m}<1)
\label{xm}
\ee
where $f$ is the fraction of layers with doping $x_p$.
Additivity of the Peltier currents implies that
the observed $\alpha$ is given by
\be
\alpha(T) = f\alpha_{p}(T) + (1-f)\alpha_{1}(T), 
\label{al}
\ee
where subscripts ${p}$ and 1 refer to quantities evaluated at $x = x_p$ and 1, 
respectively.

The scaling behavior in Fig. \ref{alpha}b implies that the second term in Eq. \ref{al} 
is negligible if $f$ is not too small.  The Peltier conductivity $\alpha_1(T)$, which has a 
maximum value of $\sim$ 0.85 A/mK, is observable only 
when $f$ becomes very small, as in the case of Sample 10* (see Supplementary Information).
Away from this limit, we may take $f \simeq 1/p$ throughout the mixed-phase region; this 
allows us to find $x_m$ using Eq. \ref{xm}.  
For e.g., in Sample 9*, the scaling of $\alpha$ shows that 1 in 40 
of the layers is conducting ($x_m$ = 0.996 instead of 1.00), 
which is consistent with the observed $\rho(300)$ in Fig. \ref{rho}a.  
In addition, we have employed other checks on the calibration.  The trilayer crystals,  
with $x = \frac34$ and $\sim$1, provide valuable calibration points.  Moreover, $x$ in Samples 1 
and 4 was determined from the $c$-axis lattice parameter measured by x-ray diffraction.  These 
checks lend support for our calibration.

%
\bfig
\incl[width=8cm]{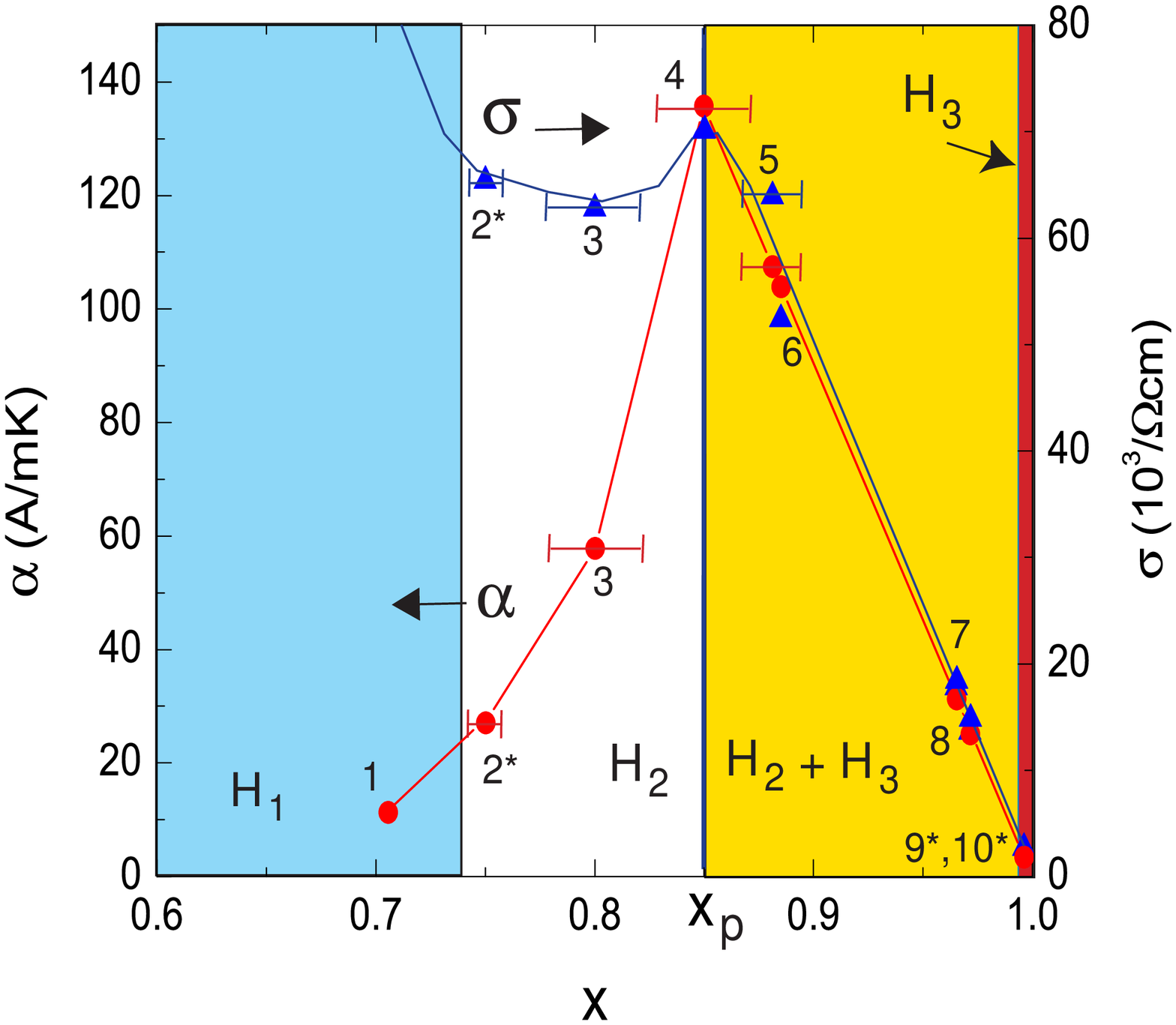}
\caption{\label{phase}
Variation of $\alpha$ (50 K) and $\sigma$ (300 K) vs. $x$ in $\rm Na_xCoO_2$ at 
large $x$ (Sample labels shown).  As $x$ traverses $H_2$, $\alpha(50)$ rises sharply to 
a maximum at $x_p$ in correlation with the increase in the CoO$_2$ layer-thickness $t$~\cite{Huang}.  
Because $\sigma(300)$ is nearly constant, the peak in $\alpha(50)$ leads to
a peak in $Z$ as well.  In $H_2+H_3$, $\alpha$ and $\sigma$ (plotted vs. $x_m$
calculated from $p$ using Eq. \ref{xm}) decrease linearly as $x_m\rightarrow 1^{-}$, implying phase-separation of holes 
into conducting layers.  The phase $H_3$ is insulating.  
In the trilayer crystals (2*, 9*, 10*) 
$x$ is locked to the commensurate values $\frac34$ and $\sim$1.  The $x$ values in Samples 1
and 4 were measured by x-ray diffraction (lines are guides to the eye).
Error bars in $x_m$ reflect the uncertainty in measuring $\rho$ and $S$ (see Supplementary Information).
}
\efig

With the inferred values of $x$ and $x_m$, we next plot the variation 
of $\alpha$ at 50 K (circles) and $\sigma$ (triangles) at 300 K in the phase diagram at large doping 
(Fig. \ref{phase}).  From $x =$ 0.71 to $x_p$, 
$\alpha(50)$ rises by a factor of 14.  The modest changes in $\sigma(300)$ (triangles) 
and $\kappa$ (see Supplementary Information) result in a substantial increase in $Z$.
We emphasize that these increases occur within the homogeneous phase $H_2$ and
are possibly correlated with the increase in layer thickness $t$~\cite{Huang}.  
Clearly, they are not a consequence of the phase separation that onsets at $x_p$.  
Both quantities decrease linearly above $x_p$.

The 40-fold enhancement of $Z$ realized at 80 K 
improves greatly the prospects for thermoelectric applications. The peak value in Fig. \ref{SvsT}b 
($Z\sim 1.8\times 10^{-3}$ K$^{-1}$) is among the highest known
for a \emph{hole}-type material below 100 K.  The promising material $\rm CsBi_4Te_6$ also 
displays a large $Z$ at 100 K~\cite{Chung}.  In comparison, $Z$ in $\rm Na_xCoO_2$ peaks at low temperatures 
whereas the curve for $\rm CsBi_4Te_6$ (curve labeled CBT) falls steeply.  
The alloy $\rm Bi_{1-y}Sb_y$ has long been known~\cite{BiSb} to 
display an even larger $Z$ at low $T$ (at 80 K, $Z = 7\times 10^{-3}$ K$^{-1}$ for $y\sim$ 0.12).  
However, it is electron-like.  To realize its advantages in a Peltier device operating below 100 K, we need
to pair Bi-Sb with a hole-type material with comparabe $Z$, but 
the 35-year search has turned up no viable candidates.  We calculate that
an optimized device with $\rm Na_xCoO_2$ ($x = x_p$) paired with $\rm Bi_{1-y}Sb_y$ 
has a device figure-of-merit $Z_{np} = 2.5\times 10^{-3}$ K$^{-1}$ at 80 K
[$Z_{np} = (S_n-S_p)^2(\sqrt{\rho_n\kappa_n}+\sqrt{\rho_p\kappa_p})^{-2}$
where subscripts n and p index the 2 materials].  This value is
higher than any reported to date.  Further enhancement of $Z$ in $\rm Na_xCoO_2$ seems
possible if we degrade $\kappa$.

{\bf Methods}\\
Sodium metal (0.5 g) and tetrahydrofuran (THF) (30 mL) in a pyrex tube were
warmed at ~100 $\rm ^o$C in an oil bath~\cite{Huang}.  Benzophenone (2 g) was added and 
the solution heated until it turned blue from
the formation of the benzophenone ketyl radical anion. 
As-grown crystals of $\rm Na_{0.75}CoO_2$ were placed along a 
pipette-like tube and immersed in the blue solution of the sodium 
radical anions. The pyrex tube was quickly capped and heated for 4 days 
at 100 $\rm ^o$C.  Finally, the crystals were extracted and washed with THF and ethanol to 
eliminate trace sodium.

This research is supported by the U.S. ONR (Contract N00014-04-1-0057) and
by NSF Grant DMR 0213706. 

Correspondence and requests for materials should be addressed to NPO (npo@princeton.edu).

Supplementary information posted online.

$^\dagger$Current address of Y.W.: Department of Physics, Univ. California, Berkeley, CA 94720.

\newpage~~
{\bf Supplementary Information}\\
\vspace{3mm}\\
\setcounter{figure}{0}
{\bf Breakdown of scaling in Sample 10*}\\
We provide more details on the measurements and analysis of the thermopower
$S$ and resistivity $\rho$ in the phase-separated region $H_2+H_3$, especially in the 
extreme case $x\rightarrow 1^{-}$ represented by Sample 10*.

\bfig[h]
\incl[width=7cm]{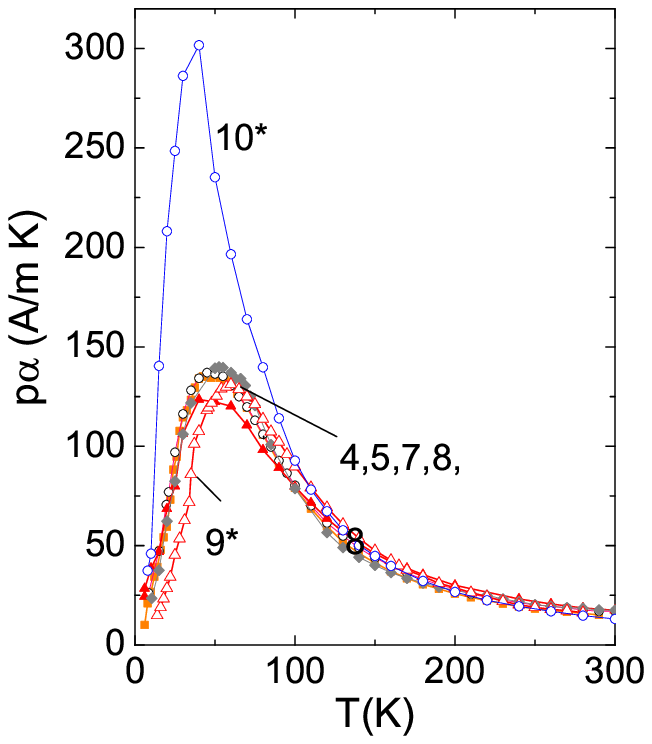}
\caption{\label{scale}
Comparison of the scaled curves of $p\alpha(T)$ in Sample 10* with those in the batch Samples
4--9* (the scale factor $p$ = 200 in Sample 10*).  Above 100 K, the scaled curve matches
those in 4--9*.  Below 100 K, the additional contribution which we identify with $\alpha_1$
in the $x = 1$ phase becomes prominent.  The peak value of $\alpha_1$ is 0.85 A/mK.  
This additional term causes the thermopower $S(T)$ in 10* to be distinct in profile from 
Samples 4--9* (see Fig. 2a of the text).
}
\efig

As noted in the text, the simplest interpretation of the scaling behavior in the curves of the Peltier conductivity 
$\alpha$ vs. $T$ is to assume that the hole carriers strongly segregate into parallel
layers throughout each crystal.  The layers must remain robustly connected electrically even
at 4 K (otherwise we would not have the metallic profile in $\rho$ with the large residual resistivity
ratios RRR$\sim$20).  Additivity of the currents implies that
the observed $\alpha$ and $\sigma$ are given by
\begin{eqnarray}
\alpha(T) &=& f\alpha_{p}(T) + (1-f)\alpha_{1}(T), \label{alpha}\\
\sigma(T) &=& f\sigma_{p}(T) + (1-f)\sigma_{1}(T), \label{sigma}
\end{eqnarray}
where subscripts ${p}$ and 1 refer to quantities evaluated at $x = x_p$ and 1, 
respectively.  

The similarity of the $\rho$-$T$ profiles in Fig. 1 of the text implies that all of $\bf J$
is carried by the conducting layers.  Setting $\sigma_1 = 0$, we have for the observed thermopower
\be
S(T) = S_p(T) + \frac{(1-f)}{f}\frac{\alpha_1(T)}{\sigma_{p}(T)},
\label{S}
\ee 
with $S_p= \alpha_p/\sigma_p$.

In Samples 5--9*, in which the curves of $\alpha$ vs. $T$ may be scaled to that in Sample 4,
the term $(1-f)\alpha_1$ must be negligible compared with
$f\alpha_{x_p}$ in Eq. \ref{alpha} (Sample 6 is excluded).  Further, if the
second term in Eq. \ref{S} is negligible (i.e. $f>$0.05), 
all the profiles of $S(T)$ should be identical to $S_p$ (Sample 4).  We find that, apart from
a 20$\%$ spread in magnitude, the curves of $S(T)$ are indeed closely similar in Samples 4--9*.

%
\bfig[h]
\incl[width=7cm]{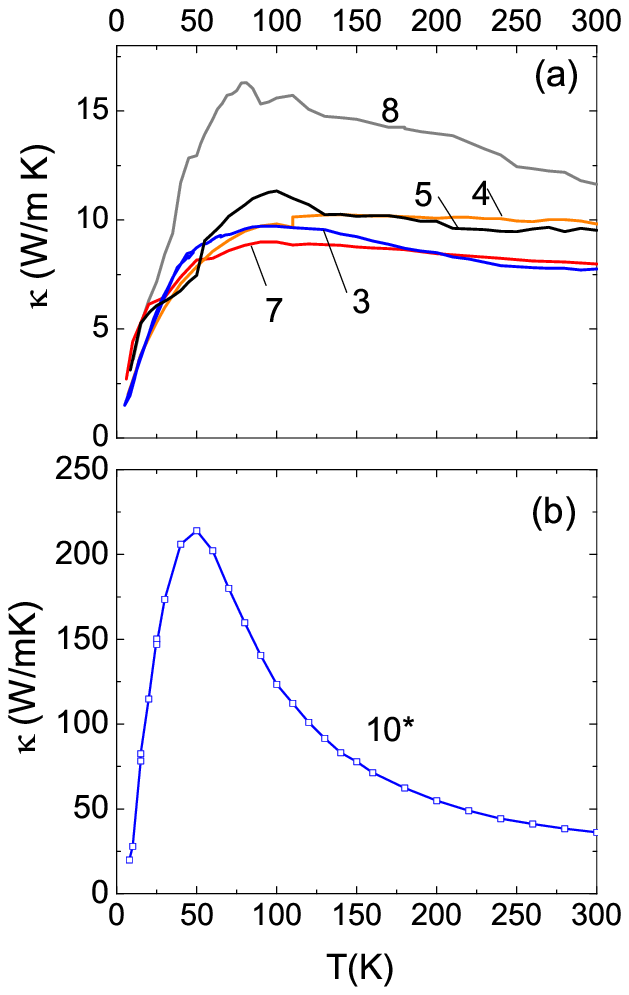}
\caption{\label{kappa}  Curves of the in-plane thermal conductivity $\kappa$ vs. $T$
in Samples 3--8 of $\rm Na_xCoO_2$ (Panel a) and in Sample 10* (Panel b). 
The magnitude of $\kappa$ = 8--10 W/mK above 50 K in Samples 3--7 is characteristic
of doped layered perovskites.  In Sample 10* ($x\simeq 1$), $\kappa$
attains much larger values, possibly reflecting the pristine nature of the Na ordering.
It is of the same order as crystals in the charge-ordered insulating state ($x$ = 0.5).
}
\efig

\bfig[h]
\incl[width=7cm]{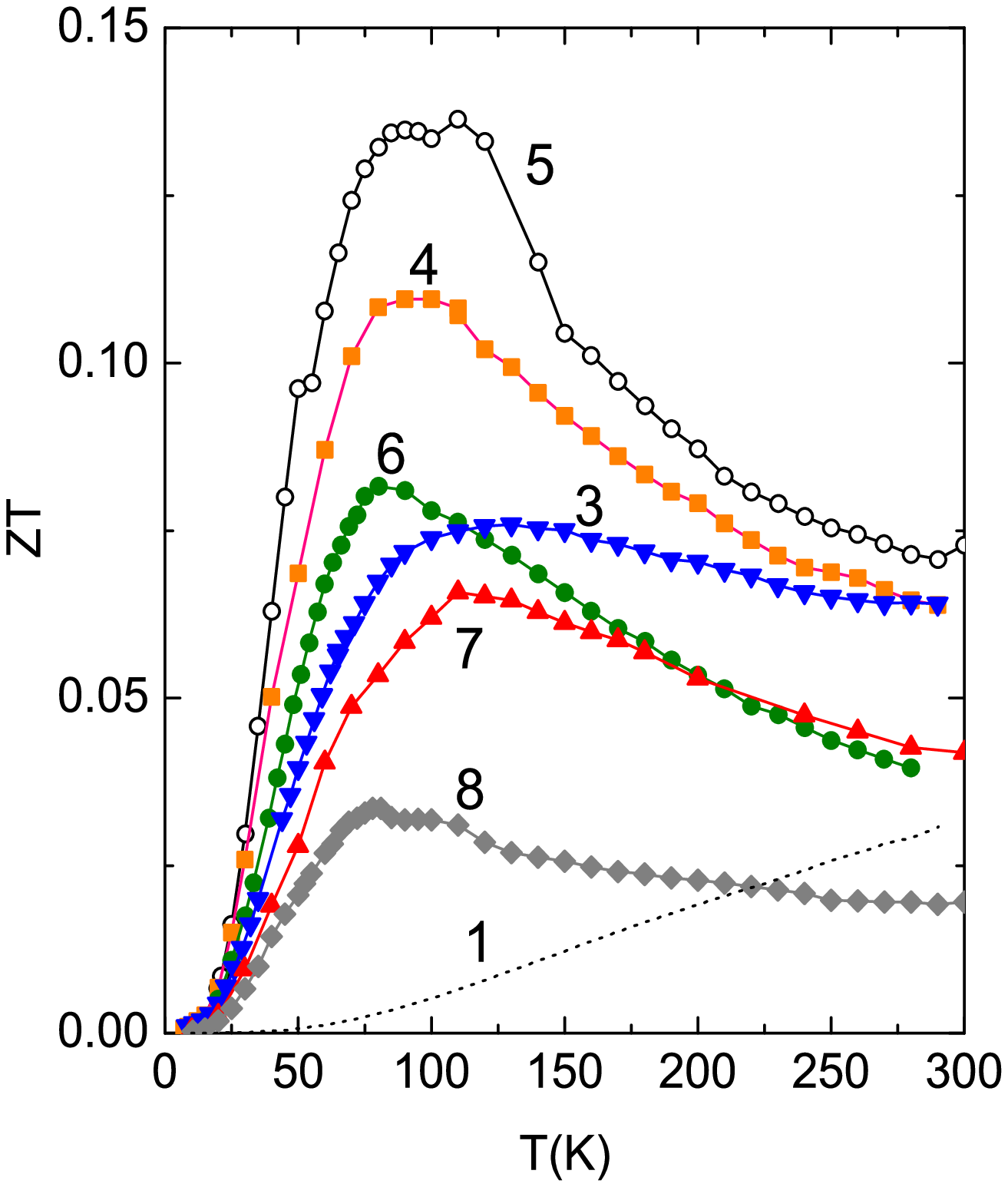}
\caption{\label{ZT}
Plots of the dimensionless figure-of-merit $ZT$ in Samples 1-8 (except 2).  
}
\efig

This spread reflects difficulties in establishing the absolute values of $S$ and $\rho$ accurately.
The size of the contact pads (50-100 $\mu$m) relative to that of the crystals (400--700 $\mu$m on a side) 
leads to uncertainties of $\pm 10\%$ in estimating the voltage lead separation.  
This is exacerbated by the large electrical anisotropy $\rho_c/\rho$ where $\rho_c$
is the $c$-axis resistivity, which causes a small
admixture of $\rho_c$ into the measured ``in-plane'' resistance despite deliberate care at
optimal contact pad placement.  We suspect that the latter factor is the main factor
that spoils exact scaling of the $\rho$-$T$ profiles.  Altogether, we estimate a combined 
uncertainty of $\pm 15\%$ in the absolute values of $\rho$.
Within this uncertainty, we see that, in Samples 4--9*, the profiles of $S$ vs. $T$ remain 
strikingly similar despite the 50-fold increase in $\rho$ in these samples.

The prominent exception is Sample 10*.  As shown in Fig. \ref{scale}, the scaled curve
$p\alpha(T)$ matches those in the other samples above 100 K.  The value of the scale factor
$p\sim$200 inferred from this plot is also 
consistent with the resistivity data ($\rho(300)$ in Sample 10*
$\sim$ 240 m$\Omega$cm nominally equals $\rho(300)$ in Sample 4 multiplied by $p$).  
Below 100 K, however,
a further contribution, which we identify with $\alpha_1(T)$ in Eq. \ref{alpha}, becomes dominant.
We note that the actual peak value of $\alpha_1$ is $\sim$ 0.85 A/mK, which is unresolvable 
in Fig. 3a of the text.  While this is a very small value, it can make the second term in Eq. \ref{alpha}
comparable to the first if $f$ becomes very small.  This is the case in Sample 10* near 50 K, for
which $f\alpha_p\sim 0.68$ A/mK.  Hence, in this extreme limit, we estimate that 1 in 200 of the layers carries the applied
current to produce a ``metallic'' resistivity profile resembling that in Sample 4 (apart from
the scale factor).  In principle, the observed thermopower and $\alpha$ should also match that of 4.
However, in Sample 10*, the term in $\alpha_1$ dominates that of the conducting layers
below 100 K and ruins the scaling behavior.

{\bf Thermal conductivity}\\
For completeness, we show the in-plane thermal conductivity measured in some of 
the samples in Fig. S2. These curves were used in computing the figure of merit $Z = S^2/\rho\kappa$, 
with $\kappa$ the in-plane thermal conductivity.  Throughout the region $H_2$,  $\kappa$ has the modest value 
10 W/mK, quite typical of layered perovskites (Fig. S2a).  However, in the limit $x$ = 1,   
$\kappa$ becomes very large (Fig. S2b), attaining values in the charge-ordered state$^1$ at $x$ = 0.5.

{\bf Dimensionless Figure of Merit $ZT$}\\
Some authors prefer to use the dimensionless figure-of-merit $ZT$ in place of $Z$.  
For comparison, we have replotted the curves of $Z$ vs. $T$ in Fig. 2b of the text as 
$ZT$ vs $T$ in Fig. S3.  An interesting feature of these curves is that $ZT$ attains a 
prominent maximum at temperatures between 60 and 100 K in the layered cobaltate.  
By comparison, in high-performance thermoelectric materials based on doped $\rm Bi_2Te_3$ 
the maximum in $ZT$ occurs in the interval 200--300 K (see for e.g. Chung \etal~$^2$).  
The rapid fall of $ZT$ in the $\rm Bi_2Te_3$  family reflects the semi-metallic nature of the bands.  
In the cobaltates, however, the peaking of $ZT$ at much lower $T$ is a consequence of the 
very narrow bands and the magnetic correlations that are present because of strong 
interaction between carriers.

{\bf References}\\
1. Foo, M. L. \etal, Phys. Rev. Lett. {\bf 92}, 247001 (2004).\\
\noindent
2. D.-Y. Chung \etal, Science {\bf 287}, 1024-1027 (2000).

\end{document}